\documentclass[
prl,
twocolumn,
superscriptaddress,
amsmath,
amssymb,
floatfix,
]{revtex4-2}

\usepackage[utf8]{inputenc}
\makeatletter
\let\latex@fnsymbol\@fnsymbol
\renewcommand\@fnsymbol[1]{\ensuremath{\ifcase#1\or * \or \dagger\or \ddagger\or
   \mathsection\or \mathparagraph\or \|\or **\or \dagger\dagger
   \or \ddagger\ddagger\or ***\or **** \else\@ctrerr\fi}}
\makeatother

\usepackage[usenames,dvipsnames]{xcolor}
\usepackage{booktabs,graphicx,mathrsfs,verbatim,soul}
\usepackage{textcomp}
\usepackage{float}
\usepackage{gensymb}
\usepackage{tabularx}
\usepackage{dcolumn} 
\usepackage[mathlines]{lineno} 
\usepackage{bm} 
\usepackage{upgreek} 
\usepackage{multirow} 
\usepackage{makecell} 
\usepackage{placeins} 
\usepackage[colorlinks=true,citecolor=blue,filecolor=blue,linkcolor=blue,urlcolor=blue]{hyperref}
\usepackage{xspace}
\usepackage[caption=false]{subfig}
\usepackage[separate-uncertainty=true]{siunitx} 
\usepackage{appendix}
\usepackage{orcidlink}
\usepackage[export]{adjustbox}

\DeclareSIUnit\clight{\text{\ensuremath{c}}}
\DeclareSIUnit\evm{\eV\per\clight\squared}
\DeclareSIUnit\year{\text{y}}
\DeclareSIUnit\day{\text{d}}
\DeclareSIUnit\tonneyears{\tonne\times\year}
\DeclareSIUnit\tonnedays{\tonne\times\day}
\DeclareSIUnit\evm{\eV\per\clight\squared}
\DeclareSIUnit\gevm{\GeV\per\clight\squared}
\DeclareSIUnit\PE{\text{PE}}
\DeclareSIUnit\ph{\text{ph}}
\DeclareSIUnit\pperkev{\ph\per\kev}
\DeclareSIUnit\events{\text{events}}
\DeclareSIUnit\pertonneyears{\per(\tonne\times\year)}
\DeclareSIUnit\mev{\mega\eV}
\DeclareSIUnit\gev{\giga\eV}

\definecolor{summer}{RGB}{86,197,150}
\definecolor{pandablue}{HTML}{203769}
\definecolor{snogreen}{HTML}{105D20}

\newcommand{\cevns}{{\text{CE}\ensuremath{\upnu}\text{NS}}\xspace}

\newcommand{\tly}{\ensuremath{t_{\mathrm{Ly}}}\xspace}
\newcommand{\tqy}{\ensuremath{t_{\mathrm{Qy}}}\xspace}

\newcommand{\ly}{\ensuremath{\mathrm{L_\mathrm{y}}}\xspace}
\newcommand{\qy}{\ensuremath{\mathrm{Q}_\mathrm{y}}\xspace}
\newcommand{\lagr}{\mathcal{L}} 

\newcommand{\beight}{\ensuremath{^8\mathrm{B} }\xspace}
\newcommand{\ybe}{\ensuremath{^{88}\mathrm{YBe}}\xspace}
\newcommand{\krcal}{\ensuremath{^\mathrm{83m}\mathrm{Kr}}\xspace}
\newcommand{\kev}{\ensuremath{\mathrm{keV}}\xspace}

\newcommand{\fluxcmsec}{\ensuremath{\mathrm{cm}^{-2}\mathrm{s}^{-1}}\xspace}
\newcommand{\Sone}{$\mathrm{S1}$\xspace}
\newcommand{\Stwo}{\ensuremath{\mathrm{S2}}\xspace}

\newcommand{\arts}{$^{37}$Ar\xspace}

\newcommand{\gevcsq}{\ensuremath{\mathrm{GeV}/c^2}}
\newcommand{\gevm}{\gevcsq}
\newcommand{\prevstwodt}{\ensuremath{\Stwo_\mathrm{pre}/\Delta t_\mathrm{pre}}\xspace}
\newcommand{\itsec}[1]{{\textit{#1} }---}

\newcommand{\rnttt}{\ensuremath{^{222}\mathrm{Rn}}\xspace}
\newcommand{\rnttz}{\ensuremath{^{220}\mathrm{Rn}}\xspace}
\newcommand{\nuis}{\theta}
\newcommand{\nuiss}{\vec{\nuis}}
\newcommand{\fref}[1]{Fig.~\ref{#1}}
\newcommand{\tref}[1]{Tab.~\ref{#1}}

\newcommand{\aref}[1]{Appendix~\ref{#1}}

\newcounter{appendix}
\renewcommand{\theappendix}{\Alph{appendix}}
\newcommand{\appendixsection}[1]{
    \refstepcounter{appendix}
    \textit{Appendix \theappendix: #1 ---} \label{app:appendix\theappendix}
}
\newcommand{\beightflux}{$(5_{-2}^{+3})\times 10^6$\,\fluxcmsec}

\newcommand{\snoflux}{$(5.25\pm0.16\mathrm{(stat.)_{-0.13}^{+0.11}\mathrm{(syst.)}})\times 10^6$\,\fluxcmsec}

\begin{document}

\title{Probing the Solar $^8$B Neutrino Fog with XENONnT}


\newcommand{\bologna}{\affiliation{Department of Physics and Astronomy, University of Bologna and INFN-Bologna, 40126 Bologna, Italy}}
\newcommand{\chicago}{\affiliation{Department of Physics, Enrico Fermi Institute \& Kavli Institute for Cosmological Physics, University of Chicago, Chicago, IL 60637, USA}}
\newcommand{\coimbra}{\affiliation{LIBPhys, Department of Physics, University of Coimbra, 3004-516 Coimbra, Portugal}}
\newcommand{\columbia}{\affiliation{Physics Department, Columbia University, New York, NY 10027, USA}}
\newcommand{\lngs}{\affiliation{INFN-Laboratori Nazionali del Gran Sasso and Gran Sasso Science Institute, 67100 L'Aquila, Italy}}
\newcommand{\mainz}{\affiliation{Institut f\"ur Physik \& Exzellenzcluster PRISMA$^{+}$, Johannes Gutenberg-Universit\"at Mainz, 55099 Mainz, Germany}}
\newcommand{\mpik}{\affiliation{Max-Planck-Institut f\"ur Kernphysik, 69117 Heidelberg, Germany}}
\newcommand{\munster}{\affiliation{Institut f\"ur Kernphysik, University of M\"unster, 48149 M\"unster, Germany}}
\newcommand{\nikhef}{\affiliation{Nikhef and the University of Amsterdam, Science Park, 1098XG Amsterdam, Netherlands}}
\newcommand{\nyuad}{\affiliation{New York University Abu Dhabi - Center for Astro, Particle and Planetary Physics, Abu Dhabi, United Arab Emirates}}
\newcommand{\purdue}{\affiliation{Department of Physics and Astronomy, Purdue University, West Lafayette, IN 47907, USA}}
\newcommand{\rice}{\affiliation{Department of Physics and Astronomy, Rice University, Houston, TX 77005, USA}}
\newcommand{\stockholm}{\affiliation{Oskar Klein Centre, Department of Physics, Stockholm University, AlbaNova, Stockholm SE-10691, Sweden}}
\newcommand{\subatech}{\affiliation{SUBATECH, IMT Atlantique, CNRS/IN2P3, Nantes Universit\'e, Nantes 44307, France}}
\newcommand{\torino}{\affiliation{INAF-Astrophysical Observatory of Torino, Department of Physics, University  of  Torino and  INFN-Torino,  10125  Torino,  Italy}}
\newcommand{\ucsd}{\affiliation{Department of Physics, University of California San Diego, La Jolla, CA 92093, USA}}
\newcommand{\wis}{\affiliation{Department of Particle Physics and Astrophysics, Weizmann Institute of Science, Rehovot 7610001, Israel}}
\newcommand{\zurich}{\affiliation{Physik-Institut, University of Z\"urich, 8057  Z\"urich, Switzerland}}
\newcommand{\paris}{\affiliation{LPNHE, Sorbonne Universit\'{e}, CNRS/IN2P3, 75005 Paris, France}}
\newcommand{\freiburg}{\affiliation{Physikalisches Institut, Universit\"at Freiburg, 79104 Freiburg, Germany}}
\newcommand{\napels}{\affiliation{Department of Physics ``Ettore Pancini'', University of Napoli and INFN-Napoli, 80126 Napoli, Italy}}
\newcommand{\nagoya}{\affiliation{Kobayashi-Maskawa Institute for the Origin of Particles and the Universe, and Institute for Space-Earth Environmental Research, Nagoya University, Furo-cho, Chikusa-ku, Nagoya, Aichi 464-8602, Japan}}
\newcommand{\laquila}{\affiliation{Department of Physics and Chemistry, University of L'Aquila, 67100 L'Aquila, Italy}}
\newcommand{\tokyo}{\affiliation{Kamioka Observatory, Institute for Cosmic Ray Research, and Kavli Institute for the Physics and Mathematics of the Universe (WPI), University of Tokyo, Higashi-Mozumi, Kamioka, Hida, Gifu 506-1205, Japan}}
\newcommand{\kobe}{\affiliation{Department of Physics, Kobe University, Kobe, Hyogo 657-8501, Japan}}
\newcommand{\kit}{\affiliation{Institute for Astroparticle Physics \& Institute of Experimental Particle Physics, Karlsruhe Institute of Technology, 76021 Karlsruhe, Germany}}
\newcommand{\tsinghua}{\affiliation{Department of Physics \& Center for High Energy Physics, Tsinghua University, Beijing 100084, P.R. China}}
\newcommand{\ferrara}{\affiliation{INFN-Ferrara and Dip. di Fisica e Scienze della Terra, Universit\`a di Ferrara, 44122 Ferrara, Italy}}
\newcommand{\groningen}{\affiliation{Nikhef and the University of Groningen, Van Swinderen Institute, 9747AG Groningen, Netherlands}}
\newcommand{\westlake}{\affiliation{Department of Physics, School of Science, Westlake University, Hangzhou 310030, P.R. China}}
\newcommand{\shenzhen}{\affiliation{School of Science and Engineering, The Chinese University of Hong Kong (Shenzhen), Shenzhen, Guangdong, 518172, P.R. China}}
\newcommand{\coimbrapoli}{\affiliation{Coimbra Polytechnic - ISEC, 3030-199 Coimbra, Portugal}}
\newcommand{\heidelberg}{\affiliation{Kirchhoff-Institute for Physics, Heidelberg University, 69120 Heidelberg, Germany}}
\newcommand{\roma}{\affiliation{INFN-Roma Tre, 00146 Roma, Italy}}
\newcommand{\bucknell}{\affiliation{Department of Physics \& Astronomy, Bucknell University, Lewisburg, PA, USA}}

\author{E.~Aprile\,\orcidlink{0000-0001-6595-7098}}\columbia
\author{J.~Aalbers\,\orcidlink{0000-0003-0030-0030}}\groningen
\author{K.~Abe\,\orcidlink{0009-0000-9620-788X}}\tokyo
\author{M.~M.~Abu~Rmeileh}\wis
\author{M.~Adrover\,\orcidlink{0123-4567-8901-2345}}\zurich
\author{S.~Ahmed~Maouloud\,\orcidlink{0000-0002-0844-4576}}\paris
\author{L.~Althueser\,\orcidlink{0000-0002-5468-4298}}\munster
\author{B.~Andrieu\,\orcidlink{0009-0002-6485-4163}}\paris
\author{E.~Angelino\,\orcidlink{0000-0002-6695-4355}}\lngs\chicago
\author{D.~Ant\'on~Martin\,\orcidlink{0000-0001-7725-5552}}\chicago
% ...
\collaboration{XENON Collaboration}\email[]{xenon@lngs.infn.it}\noaffiliation


\date{\today}

\begin{abstract}
We report a 3.3\,$\sigma$ measurement of coherent elastic neutrino--nucleus scattering from solar $^8$B neutrinos using a 6.77\,t$\times$yr exposure from the XENONnT experiment, inferring a solar $^8$B neutrino flux of $(5_{-2}^{+3})\times 10^6\,\mathrm{cm}^{-2}\mathrm{s}^{-1}$, consistent with previous measurements. In the presence of the $^8$B ``neutrino fog'', we find no evidence for light dark matter, and observe diminishing returns in sensitivity with increasing exposure. A 93\% increase in exposure from the previous search improves the median sensitivity to the nucleon scattering cross section for a 5\,GeV/$c^2$ spin-independent weakly interacting massive particle by 10\%. The dataset was also used to measure the weak mixing angle at $\sim$ 0.02\,GeV/$c$ momentum transfer and constrain physics beyond the Standard Model.
\end{abstract}

\maketitle

\itsec{Introduction}
Solar \beight neutrinos from the proton--proton chain were first detected by the Homestake experiment in 1968~\cite{Davis:1968cp}. Precision measurements by Super-Kamiokande~\cite{Super-Kamiokande:1998qwk}, the Sudbury Neutrino Observatory\,(SNO)~\cite{SNO:2011hxd}, and Borexino~\cite{Borexino:2008fkj} established solar neutrinos as a cornerstone of neutrino physics and enabled stringent tests of the Standard Solar Model~\cite{Bahcall:1996qv}. Coherent elastic neutrino--nucleus scattering\,(\cevns)~\cite{Freedman:1973yd} is experimentally challenging to observe: the low momentum transfer involved requires both a low energy threshold and a stringently suppressed background~\cite{COHERENT:2017ipa, Ackermann:2025obx}. Owing to a cross section that scales approximately with the square of the neutron number, the \cevns process induced by solar \beight neutrinos has only recently been detected by multi-ton liquid-xenon\,(LXe) dark matter\,(DM) detectors: XENONnT~\cite{XENON:2024ijk} and PandaX-4T~\cite{PandaX:2024muv}.

Beyond its intrinsic interest as a neutrino process, \beight~\cevns probes neutrino interactions at low momentum transfer and provides sensitivity to physics beyond the Standard Model\,(BSM)~\cite{AtzoriCorona:2025ygn}. The nuclear recoil\,(NR) signature of \beight~\cevns closely resembles that of $\mathcal{O}(5)\,\mathrm{GeV}/c^2$ weakly interacting massive particles\,(WIMPs)~\cite{XENON:2024hup,XENON:2025vwd}. In this Letter, we report a measurement of \beight~\cevns, present search results for WIMPs in the ``neutrino fog''~\cite{Billard:2013qya,OHare:2021utq}, and investigate additional BSM neutrino physics channels.

\itsec{Experiment}
Located at the INFN Laboratori Nazionali del Gran Sasso in Italy, the XENONnT experiment~\cite{XENON:2024wpa} searches for NRs induced by WIMPs scattering off xenon nuclei. The main detector is a dual-phase time projection chamber\,(TPC). The active target consists of 5.9\,t of LXe cylindrically enclosed by polytetrafluoroethylene\,(PTFE) walls and monitored by 494 3-inch Hamamatsu R11410-21 photomultiplier tubes\,(PMTs)~\cite{Antochi:2021wik} arranged in top and bottom arrays. The TPC is housed inside a double-walled cryostat, surrounded by a water-based, gadolinium-loaded neutron veto, and an outer muon veto~\cite{XENON:2024fxf, XENON1T:2014eqx}. 

Particle interactions in the LXe produce prompt scintillation photons and liberated electrons. The prompt photons are referred to as S1 signals. The freed electrons drift upward under an applied electric field to the liquid--gas interface, where a stronger electric field extracts them into the gaseous phase. There, they generate a secondary proportional scintillation signal referred to as the S2 signal. The time difference between S1 and S2 gives the vertical interaction depth\,(Z), while the S2 hit pattern in the top PMT array determines the position in the horizontal XY-plane. The S1 and S2 amplitudes and their ratio allow reconstruction of the deposited energy and discrimination of NR from electronic recoil\,(ER).

\itsec{Datasets}
This search combines three datasets with a total live time of 603 days. The unchanged datasets from the first two science runs\,(SR0 and SR1) provide 316.5 live days after correcting for data-acquisition dead time and vetoes~\cite{XENON:2024ijk}. A third dataset\,(SR2) was acquired between October~12,~2023 and March~25,~2025, contributing an additional 286.5 live days. During SR2, the temperature and pressure in the TPC were stable, with an average of $(178.4 \pm 0.5)$\,K and $(2.038 \pm 0.017)$\,bar. The liquid level was at $(4.83 \pm 0.01)$\,mm above the gate throughout the SR2. The drift electric field remained at \SI{23}{V/cm}. Comparing to the previous SRs, in SR2 a pressure increase of \SI{0.1}{bar} (corresponding to \SI{1}{K} higher temperature) caused the drift velocity to increase from $(0.675 \pm 0.006)$\,mm/$\upmu$s to $(0.691 \pm 0.005)$\,mm/$\upmu$s. Consequently, the maximum drift time\,($t_\mathrm{drift}^{\max}$) decreased from $(2.20 \pm 0.01)$\,ms to $(2.18 \pm 0.01)$\,ms. The anode voltage was also reduced from 4.95\,kV to 4.85\,kV, resulting in a slightly lower extraction field.

In addition to the 20 PMTs already excluded from the analysis in SR0 and SR1, one additional PMT was disabled due to intermittent light emission. PMT gains were monitored weekly using pulsed LED signals, with 85\% of PMTs stable within 2.5\%. Hits were recorded as a PMT triggered the digitization threshold, typically about \SI{2.06}{mV}~\cite{XENON:2022vye}. The mean single photoelectron\,(PE) acceptance, determined from LED calibration data, was $(92.6 \pm 0.1)\%$. The PMT hits are clustered into peaks based on their timing and subsequently classified as S1 or S2 according to their waveform using a self-organizing maps\,(SOM)~\cite{Kohonen1982} method. For S2 signals, an iterative Bayesian algorithm implemented in \textsc{strax}~\cite{strax} and \textsc{straxen}~\cite{straxen} evaluates peak time and position to prevent the merging of multiple-site interactions.

Distortions of the drift field near the edges of the detector lead to discrepancies between the true interaction position (X, Y) and the position (X$_\text{rec}$, Y$_\text{rec}$) reconstructed from the top PMT hit pattern. A data-driven radial correction is derived from \krcal calibration data, with positions corrected relative to a boundary defined by the simulated drift field to account for the charge-insensitive volume~\cite{XENONnT:2023dvq}. Additionally, a field-dependent drift velocity is used to correct the Z coordinate. The cylindrical fiducial volume\,(FV) is defined as radius $\mathrm{R} < \SI{60.1}{cm}$ and $\SI{-142}{cm} < \mathrm{Z} < \SI{-13}{cm}$, restricting the analysis to a region where the detector response and background modeling are well understood. The resulting mass of FV is $(4.14 \pm 0.14)$\,t, with a 3.3\% uncertainty originating from position reconstruction resolution and drift field modeling. The total exposure of this analysis is 6.77\,t$\times$yr.

Light from S1 or S2 signals can generate delayed electron signals via photoionization of impurities in the LXe~\cite{XENON:2021qze}. The photoionization strength is defined as the ratio between the number of photoionization electrons detected within $t_\mathrm{drift}^{\max}$ after an S2 signal and the number of electrons in that S2 signal itself. Following a maintenance period between SR0 and SR1, the photoionization strength increased by approximately an order of magnitude. During SR2, the installation of a hydrocarbon filter resulted in a subsequent decrease in the photoionization rate. Position- and time-dependent signal inhomogeneities are corrected~\cite{XENON:2022ltv}. The month-scale time variation of the photoionization strength in SR2 necessitates additional time-dependent corrections to the S1 and S2 signal integrated waveform areas. The ``electron lifetime'', representing the attenuation of drifting ionization electrons due to electronegative impurities, is modeled using \rnttt and \krcal calibration data~\cite{XENON:2024mlv} and measured to be above 20\,ms in SR2.

After all corrections, the relative temporal stability of the light and charge yields\,(\ly and \qy) in SR2 is within 0.35\% and 2.55\%, respectively. These variations are propagated as systematic uncertainties in the determination of the photon and electron gains, $g_1$ and $g_2$. Using the method described in Ref.~\cite{XENON:2022ltv}, the values of $g_1$ and $g_2$ in SR2 are measured to be $(0.1378 \pm 0.0010)\,\mathrm{PE/photon}$ and $(14.3 \pm 0.4)\,\mathrm{PE/electron}$. The value of $g_2$ is lower than that in SR1, $(16.9 \pm 0.5)\,\mathrm{PE/electron}$~\cite{XENON:2024ijk}, resulting in reduced S2 signals.

\begin{figure}[!t]
    \centering
    \includegraphics[width=1.0\columnwidth,left]{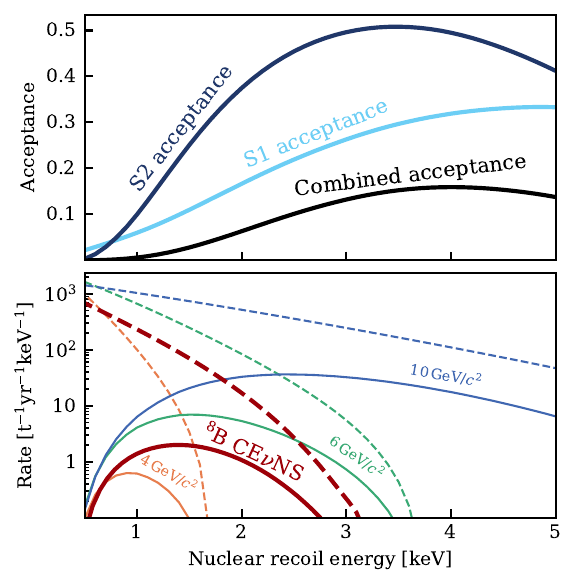}
    \caption{Top: acceptance of low-energy NR, averaged over the three SRs weighted by exposure. The light\,(dark) blue curve shows the S1\,(S2) detection acceptance, and the black curve shows the combined acceptance. Bottom: energy spectra of \beight~\cevns and spin-independent\,(SI) WIMPs. Solid\,(dashed) lines show spectra with\,(without) acceptance. An SI WIMP--nucleon cross section of $10^{-44}$\,cm$^2$ is assumed for the WIMP spectra.}
    \label{fig:detection_acceptance}
\end{figure}

\itsec{\cevns and LDM Signal}
The expected NR spectrum from \beight~\cevns is shown in \fref{fig:detection_acceptance}. It is computed using the \beight neutrino energy spectrum given by the Standard Solar Model~\cite{Bahcall:1996qv} and the Standard Model\,(SM) \cevns cross section on xenon nuclei~\cite{Barranco:2005yy}, accounting for the natural isotopic abundances. At tree level, the differential cross section in terms of the NR energy $T$ is given by $\mathrm{d}\sigma/\mathrm{d}T \propto G_V^2$, where the vector coupling $G_V \propto g_V^p Z + g_V^n N$ depends on the proton\,(neutron) number $Z$\,($N$) and the weak mixing angle $\theta_W$, with $g_V^p = \frac{1}{2} - 2\sin^2\theta_W$ and $g_V^n = -\frac{1}{2}$.

The low-energy NR \ly and \qy were calibrated in SR1 using a dedicated \ybe calibration source~\cite{XENON:2024kbh}. The corresponding uncertainties are modeled by nuisance parameters, $\tly$ and $\tqy$, defined such that $\tly=0$\,($\tqy=0$) corresponds to the median of \ly\,(\qy), while $\tly=\pm1$\,($\tqy=\pm1$) corresponds to the $\pm1\sigma$ quantiles.

The NR yields in all SRs are updated relative to those used in Ref.~\cite{XENON:2024ijk}. This revision is primarily driven by improved modeling of the S1 reconstruction efficiency for neutron multiple-scattering events in the \ybe calibration dataset. At \SI{1.5}{keV}, the median light\,(charge) yield is reduced by 7\%\,(10\%). The downward shift of the median yields leads to an overall 25\% reduction in the total number of expected \beight events. For NR energies outside [0.5,\,5.0]\,keV, where the \ybe calibration does not provide reliable yield measurements, the yields are set to zero.

Because of the low-energy nature of the recoils, S1 signals with two or three triggered PMTs and S2 signals between 120 and 500\,PE (corresponding to about 4 and 18 extracted electrons) are selected as the region of interest\,(ROI) in all SRs. To suppress random PMT pileup, S1 signals must have at least two hits within $\pm50$\,ns of the maximum of the waveform. The S1 detection efficiency is evaluated using the waveform simulation framework \textsc{fuse}~\cite{fuse} and validated by comparing to multiple calibration datasets. The S2 detection efficiency above \SI{120}{PE} is unity, as verified by waveform simulations. Events are reconstructed as in Ref.~\cite{XENON:2024mlv}, and the event-building efficiency is evaluated by injecting simulated events into real data and processing the combined stream with \textsc{axidence}~\cite{axidence}.

Events in ROI are blinded before the full analysis procedure is finalized. In addition to the ROI selection, quality cuts are applied to ensure well-reconstructed events~\cite{XENON:2024ijk} while maintaining high acceptance, calculated using data near the cathode within the ROI. The gate and anode electrodes are supported by additional perpendicular wires, introducing distortions in the S2 waveform shape that are insufficiently constrained for precise modeling. These S2 signals are therefore removed using a wire boundary that depends on the S2 area and accounts for the (X, Y) position reconstruction resolution.

The S1, S2, and combined detection acceptances as functions of NR energy, together with the \cevns and spin-independent\,(SI) WIMPs spectra, are shown in \fref{fig:detection_acceptance}. Among detectable \beight~\cevns events, 90\% have recoil energies between 0.75 and 2.25\,keV and neutrino energies between 9.2 and 13.8\,MeV. The S1 acceptance is primarily determined by the S1 ROI selection. The S2 acceptance includes the S2 ROI requirement, event selection efficiency, and event-building efficiency. It further incorporates additional rejection criteria to suppress the accidental coincidence\,(AC) background, as described in the background section. The \beight signal rate is lower in SR2 compared to SR1 due to the reduced $g_2$. The relative uncertainty of S1 and S2 acceptance is 15\%.

Elastic NR light dark matter\,(LDM) can produce recoil spectra similar to that of \beight~\cevns for particle masses in the range 3--12\,\gevcsq. A spin-independent\,(SI) DM--nucleon interaction~\cite{Lewin:1995rx}, mediated by a heavy mediator with mass $m_\phi \gg q_0/c \equiv \SI{20}{\mev}/c^2$ is considered, where $q_0$ denotes the typical momentum transfer in this search. The corresponding feature distributions are simulated with the NR spectrum appropriately modified according to the interaction hypothesis.

\begingroup
\renewcommand{\arraystretch}{1.3}
\begin{table}[!t]
    \centering
    \caption{Expected and best-fit event numbers for the signal and background components within the ROI, where the \beight~\cevns rate is left unconstrained in the fit. The uncertainty on the signal expectation includes contributions from the combined acceptances and (\ly, \qy). The quoted uncertainties for the background components reflect the widths of constraints applied in the fit. The total number of best-fit events differs from the observed number of events because the best-fit rates are determined by both the event distributions and the auxiliary terms. \label{tab:backgrounds}}
    \vspace{0.1cm}
    \begin{tabular}{ccc}
    \hline\hline
    Component & Expectation & Best-fit \\
    \hline
    AC (SR0)  & $7.5 \pm 0.7$ & $7.4 \pm 0.7$ \\
    AC (SR1)  & $17.8 \pm 1.0$ & $17.9 \pm 1.0$ \\
    AC (SR2)  & $14.9 \pm 0.7$ & $14.9 \pm 0.7$ \\
    ER        & $1.2 \pm 1.2$ & $1.4 \pm 1.1$ \\
    Neutron   & $0.8^{+0.4}_{-0.5}$ & $0.8 \pm 0.4$ \\
    \hline
    Total background & $42.1^{+1.9}_{-1.7}$ & $42.4 \pm 1.9$ \\
    \hline
    \beight~(SR0) & $3.2^{+1.0}_{-0.9}$ & $3.3 \pm 1.0$ \\
    \beight~(SR1) & $5.9^{+1.9}_{-1.6}$ & $6.1^{+1.7}_{-1.9}$ \\
    \beight~(SR2) & $7.2^{+2.3}_{-1.9}$ & $7.3^{+2.1}_{-2.3}$ \\
    \hline
    Total \beight & $16^{+5}_{-4}$ & $17 \pm 5$ \\
    \hline
    Total observed & \multicolumn{2}{c}{9 (SR0) + 28 (SR1) + 25 (SR2) = 62} \\
    \hline\hline
    \end{tabular}
\end{table}
\endgroup

\itsec{Backgrounds}
This search for solar \beight~\cevns considers AC, neutron, ER, and surface background components, using the same modeling procedure for each as in the first XENONnT analysis~\cite{XENON:2024ijk}. In the search for LDM, the \beight~\cevns signal is considered as a background. The expected signal and background event numbers in the ROI are listed in \tref{tab:backgrounds}.

The background in the ROI is dominated by AC events, arising from accidental pairing of ``isolated'' \Sone and \Stwo signals. The primary source of these isolated signals is delayed signals caused by high-energy\,(HE) interactions, primarily due to $\mathcal{O}(1)$\,MeV $\gamma$ rays from materials' radioactivity. 
In the previous search~\cite{XENON:2024ijk}, variables quantifying the temporal and spatial correlation with preceding HE events were used to suppress AC background, e.g. \prevstwodt, where $\Stwo_\mathrm{pre}$ is the \Stwo area of the preceding HE event and $\Delta t_\mathrm{pre}$ is the time separation between that event and the isolated signal.
In SR2, these quantities are incorporated into event building: event-triggering \Stwo signals are first selected using these quantities, and \Sone signals within $t_\mathrm{drift}^{\max}$ are associated to form events. This procedure improves the temporal stability and spatial uniformity of the isolated S2 rate arising from HE-induced photoionization and localized single electron bursts. \tref{tab:iso_rate} summarizes the isolated signal rates and predicted AC rates in the three SRs. In SR2, the isolated \Sone and \Stwo rates are both reduced by about 20\% compared to SR1, following improvements in peak classification and event reconstruction.

\begin{table}[!t]
  \centering
  \caption{Isolated \Sone, \Stwo signals and AC event rate in the three SRs. The AC event rate is after all event selections.}
  \vspace{0.1cm}
  \begin{tabular}{cccc}
    \toprule
    SR & {S1\,[\unit{Hz}]} & { S2\,[\unit{mHz}]} & { AC\,[$\upmu$Hz]} \\
    \midrule
    SR0	& 2.3 & 16.6 & 0.80 \\
    SR1	& 2.2 & 21.5 & 0.99 \\
    SR2 & 1.7 & 17.7 & 0.60 \\
    \bottomrule
  \end{tabular}
  \label{tab:iso_rate}
\end{table}


The AC background is modeled by the same data-driven modeling method used in the previous search~\cite{XENON:2024ijk}. Two boosted decision tree\,(BDT) classifiers are trained on S1 and S2 waveform features to discriminate between \beight~\cevns from AC events, following Ref.~\cite{XENON:2024ijk}. A selection based on S2 BDT score at 80\% \beight~\cevns signal acceptance reduces the expected AC event count in SR2 from 432.3 to 14.9. The relative systematic uncertainty of the acceptance is evaluated to be 10\%, estimated from the deviation between simulation and data in the ROI near the cathode and PTFE wall because no other data in a similar ROI are available. The events rejected by the S2 BDT selection are used as a sideband dataset. A total of 441 events were observed in the sideband, in agreement with the predicted 417.4 events. In addition, the distribution of events in the analysis space is consistent with the model predictions, as detailed in the statistical inference section. The sideband's relative statistical uncertainty of 4.9\% is therefore used to estimate the uncertainty on the predicted AC rate in the science dataset.

Surface events arising from plated-out \rnttt daughters on the PTFE wall can leak into the ROI~\cite{XENON:2024xgd}. Their radial distribution is modeled, incorporating the S2-dependent position reconstruction uncertainty. A selection of events with $\mathrm{R} < \SI{60.1}{cm}$ suppresses the residual surface background to 0.1\,/(t$\times$yr), rendering it negligible for this analysis, as demonstrated by a toy Monte Carlo\,(MC) study.

The neutron contribution is evaluated with a full simulation chain, using neutron spectra derived from material radioactivity, updated low-energy NR yields, and an improved neutron veto tagging efficiency of 76\% because of gadolinium-doping in SR2. This results in a prediction of $0.8^{+0.4}_{-0.5}$ events for the three SRs combined, where the uncertainty reflects the NR rate and yield modeling.

The ER background is modeled with a flat spectrum and normalized using rates directly from online data-quality monitoring, given the large uncertainty tolerance. To account for discrepancies in the ER emission modeling between fits to \rnttz~\cite{XENON:2024xgd} and \arts calibration data under the NEST parameterization~\cite{Szydagis:2022ikv}, a conservative 100\% uncertainty is assigned. The total expected ER background across the three SRs is $1.2 \pm 1.2$ events.

\itsec{Statistical Inference}
Corrected S2 signal amplitude\,(cS2), \prevstwodt, S1 BDT score, and S2 BDT score are the observables in this analysis and used to discriminate between the \beight~\cevns or LDM signal and backgrounds. The data are partitioned into three bins per observable, resulting in 81 bins in the four-dimensional parameter space, and an extended binned likelihood fit is performed. The binning is defined such that, in the projection onto each dimension, the expected number of AC events per bin is the same. The limited number of bins and the uniform AC population are adopted to reduce potential biases associated with the finite statistics of isolated peaks. This choice is validated with toy MC simulations.

\begin{figure}[!t]
    \centering
    \includegraphics[width=0.95\columnwidth,left]{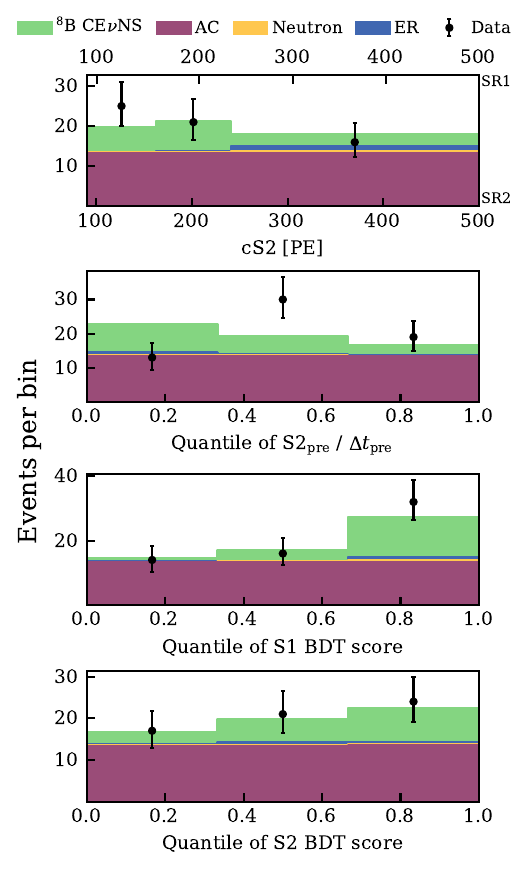}
    \caption{Best-fit signal and background distributions compared to data in the projected analysis dimensions, combining all SRs. Black markers show the observed event counts with Poisson uncertainties. The stacked histograms display the \beight~\cevns signal\,(light green) and background components: AC\,(purple), ER\,(blue), and neutrons\,(yellow). Due to SR-dependent binning, cS2 is shown with a double axis, while the other dimensions are presented in AC quantiles for the summed dataset.}
    \label{fig:best_fit}
\end{figure}

The likelihood function is defined as $\lagr(\mu,\nuiss) = \prod_{i=0,1,2} \lagr_{i}(\mu,\nuiss)\times \prod_m \lagr_m(\nuis_m)$, where $\mu$ denotes the parameter of interest. The set $\nuiss$ represents nuisance parameters constrained by Gaussian auxiliary terms. The index $i$ runs over the three SRs, while $m$ labels the individual nuisance contributions. The nuisance terms include uncertainties in the AC, neutron, and ER background rates, the signal acceptance, and the NR yield model parameters (\tly, \tqy). The AC background rates, which dominate the total background, are treated as independent between SRs; all remaining background components are constrained with correlated rates.

The \beight~\cevns discovery significance, the constraint on the \beight neutrino flux, limits on the LDM cross section, and bounds on BSM neutrino parameters are derived using the profile log-likelihood ratio test statistic, following Ref.~\cite{XENON:2024xgd}. The expected discovery significance and the critical regions for confidence interval construction are determined with toy MC simulations implemented in \textsc{alea}~\cite{alea}. The agreement between the best-fit model and the data is assessed through goodness-of-fit\,(GOF) tests based on the binned likelihood. Four tests are performed on the one-dimensional projections, and an additional test is carried out in the full four-dimensional space. The corresponding $p$-values are obtained from the distributions of the GOF test statistics generated with toy MC simulations. A threshold of 0.010 is applied to each independent test, corresponding to a combined 95\% confidence level\,(CL) requirement. The test definition is fixed before unblinding, and its sensitivity to potential mis-modeling is validated with toy MC.

\begin{figure}[!t]
    \centering
    \includegraphics[width=1.0\columnwidth,left]{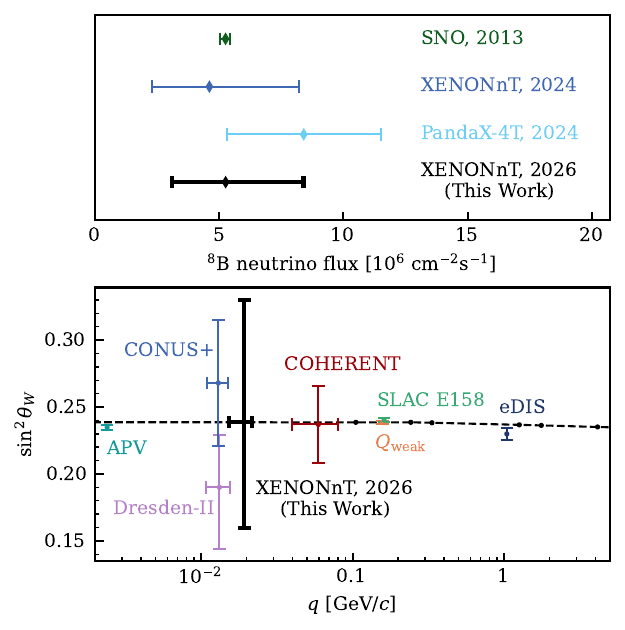}
    \caption{Constraints on the solar \beight neutrino flux and the weak mixing angle $\theta_W$. Top: the 68\% confidence interval on the solar \beight flux obtained from the first three SRs of XENONnT, compared with measurements from SNO (\snoflux)~\cite{SNO:2011hxd}, the first two SRs of XENONnT~\cite{XENON:2024ijk}, and PandaX-4T~\cite{PandaX:2024muv}. Bottom: the extracted value of $\sin^2\theta_W$ assuming the SNO \beight flux~\cite{SNO:2011hxd}, shown alongside previous determinations~\cite{ParticleDataGroup:2024cfk,Wood:1997zq,Qweak:2018tjf,SLACE158:2005uay,Prescott:1979dh,Wang:2014guo,AristizabalSierra:2022axl,DeRomeri:2022twg,Alpizar-Venegas:2025wor} and the SM prediction in the modified minimal subtraction\,($\overline{\text{MS}}$) renormalization scheme~\cite{Erler:2017knj}.}
    \label{fig:weinberg_angle_comparison}
\end{figure}

Motivated by the low $p$-value observed in the one-dimensional \prevstwodt GOF test in the combined SR0 and SR1 analysis~\cite{XENON:2024ijk}, an additional SR2-only GOF test on \prevstwodt was planned to be performed immediately after unblinding the SR2 data. If the resulting $p$-value was below 0.05, the final inference would be done in three dimensions (cS2, S1 BDT score, S2 BDT score), excluding \prevstwodt. The distribution of the test statistic under the background-only hypothesis is modeled using the asymptotic approximation~\cite{Chernoff:1954eli,Cowan:2010js}. This conditional treatment of \prevstwodt is verified not to compromise the validity of the asymptotic approximation used for the discovery significance.

\begin{figure}[!t]
    \centering
    \includegraphics[width=1.0\columnwidth,left]{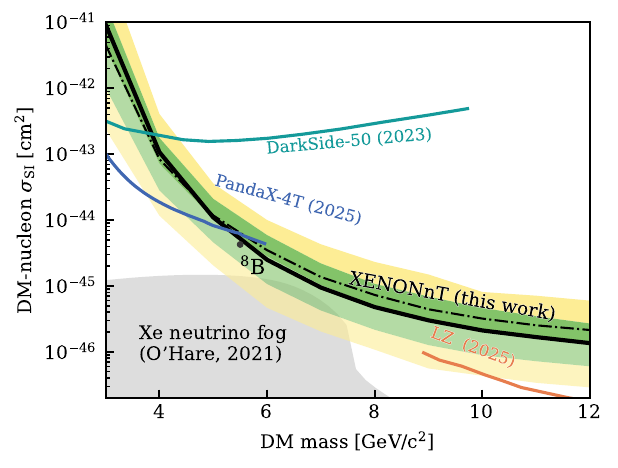}
    \caption{Upper limit on SI WIMP--nucleon interaction at 90\% CL. The green and yellow bands represent the 1$\sigma$ and 2$\sigma$ sensitivity expectations, respectively. The black solid line shows the result of this work, which incorporates updated yields and demonstrates improved sensitivity. The black dash-dotted line shows the SR0+1 sensitivity median, re-evaluated using the same updated yields. Results from other experiments~\cite{PandaX:2025rrz, LZ:2024zvo, DarkSide-50:2023fcw} are also shown. The neutrino fog defined in Ref.~\cite{OHare:2021utq} is shown in the gray shaded region. A \beight~\cevns equivalent cross section for a $\SI{5.5}{\gev}/c^2$ WIMP is $4.3\times10^{-45}$\,\unit{cm^2}.}
    \label{fig:light_dark_matter_limits_wimp_si}
\end{figure}

\itsec{Results}
After unblinding, 25 events are observed in the SR2 dataset as shown in \fref{fig:best_fit}. Both the event count and their distribution are consistent with the expectation from \beight~\cevns on top of background contributions. A fit to SR2 data alone results in a $p$-value of 0.43 in the one-dimensional GOF test on \prevstwodt. Consequently, the four-dimensional binned likelihood analysis is applied to the combined dataset.

Including the 9 and 28 events in SR0 and SR1, respectively, the total number of observed events is 62. The projections of the best-fit model in the four analysis dimensions, together with the binned event distributions, are shown in \fref{fig:best_fit}; the individual SR projections are provided in \aref{app:appendixA}. All best-fit nuisance parameters $\nuiss$ lie within $\pm0.2\sigma$ of their external Gaussian constraints. The five GOF tests of the best-fit model provide $p$-values above the predefined threshold, except for the one-dimensional \prevstwodt test ($p=0.008$). This behavior is expected given the previously observed tension in the SR0 and SR1 datasets~\cite{XENON:2024ijk}. The background-only hypothesis is rejected with a discovery significance of $3.3\sigma$, corresponding to a $p$-value of $4.5\times10^{-4}$.

Assuming the SM prediction for the \cevns cross section, the resulting solar \beight neutrino flux is \beightflux, as shown in the top panel of \fref{fig:weinberg_angle_comparison}. The new constraint is consistent within uncertainties with that obtained from the first two SRs using a different NR yield model~\cite{XENON:2024ijk}.

No excess beyond the \beight~\cevns and backgrounds is observed. Taking the solar \beight neutrino flux from SNO constraints~\cite{SNO:2011hxd}, the upper limit on the SI WIMP--nucleon interaction cross section is therefore set at 90\% CL, as shown in \fref{fig:light_dark_matter_limits_wimp_si}. As experiments approach the ``neutrino fog''~\cite{Billard:2013qya, OHare:2021utq}, further increases in exposure lead to diminishing returns. Relative to the previous search~\cite{XENON:2024ijk}, a 93\% increase in exposure improves the median sensitivity to a $\SI{5}{\gev}/c^2$ SI WIMP by only 10\%. The  $\pm1\sigma$ interval for $\sin^2\theta_W$ is $[0.159,\,0.330]$ at $\sim0.02\,\mathrm{GeV}/c$ momentum transfer, as shown in the bottom panel of \fref{fig:weinberg_angle_comparison}.

\itsec{Summary and Outlook}
A blind search for NRs induced by solar \beight neutrinos via \cevns has been carried out with XENONnT using three SRs, corresponding to a total exposure of 6.77\,t$\times$yr. Dedicated techniques are developed to suppress the dominant AC background. The combined dataset rejects the background-only hypothesis at a significance of 3.3\,$\sigma$. The inferred solar \beight neutrino flux, \beightflux, is consistent with previous measurements.

The same dataset is further used to constrain LDM--nucleon interaction cross section through elastic NRs, illustrating the diminishing returns of increased exposure in the presence of the ``neutrino fog''. In addition, it probes BSM neutrino interactions, including the weak mixing angle at low momentum transfer, new vector mediators, and a finite neutrino charge radius\,(presented in the End Matter). These results demonstrate the versatility of the LXe TPC as a precision instrument for low-energy neutrino measurements and for searches for new physics.

\itsec{Acknowledgements}
We gratefully acknowledge support from the National Science Foundation, Swiss National Science Foundation, German Ministry for Education and Research, Max Planck Gesellschaft, Deutsche Forschungsgemeinschaft, Helmholtz Association, Dutch Research Council (NWO), Fundacao para a Ciencia e Tecnologia, Weizmann Institute of Science, Binational Science Foundation, Région des Pays de la Loire, Knut and Alice Wallenberg Foundation, Kavli Foundation, JSPS Kakenhi, JST FOREST Program, and ERAN in Japan, Tsinghua University Initiative Scientific Research Program, National Natural Science Foundation of China, Ministry of Education of China, DIM-ACAV+ Région Ile-de-France, and Istituto Nazionale di Fisica Nucleare. This project has received funding/support from the European Union’s Horizon 2020 and Horizon Europe research and innovation programs under the Marie Skłodowska-Curie grant agreements No 860881-HIDDeN and No 101081465-AUFRANDE.

We gratefully acknowledge support for providing computing and data-processing resources of the Open Science Pool and the European Grid Initiative, at the following computing centers: the CNRS/IN2P3 (Lyon - France), the Dutch national e-infrastructure with the support of SURF Cooperative, the Nikhef Data-Processing Facility (Amsterdam - Netherlands), the INFN-CNAF (Bologna - Italy), the San Diego Supercomputer Center (San Diego - USA) and the Enrico Fermi Institute (Chicago - USA). We acknowledge the support of the Research Computing Center (RCC) at The University of Chicago for providing computing resources for data analysis.

We thank the INFN Laboratori Nazionali del Gran Sasso for hosting and supporting the XENON project.

\itsec{Data Availability}
The data and components distributions that support this article are openly available~\cite{release}.

\vspace{0.05cm}

\itsec{Note Added}
We note the recent preprint from the LZ Collaboration reporting on a search for \beight neutrinos and LDM~\cite{LZ:2025igz}.

\bibliography{bibliography}

\onecolumngrid
\begin{center}
{\large\bfseries End Matter}
\end{center}
\begin{figure*}[!h]
  \centering
  \includegraphics[width=0.85\textwidth]{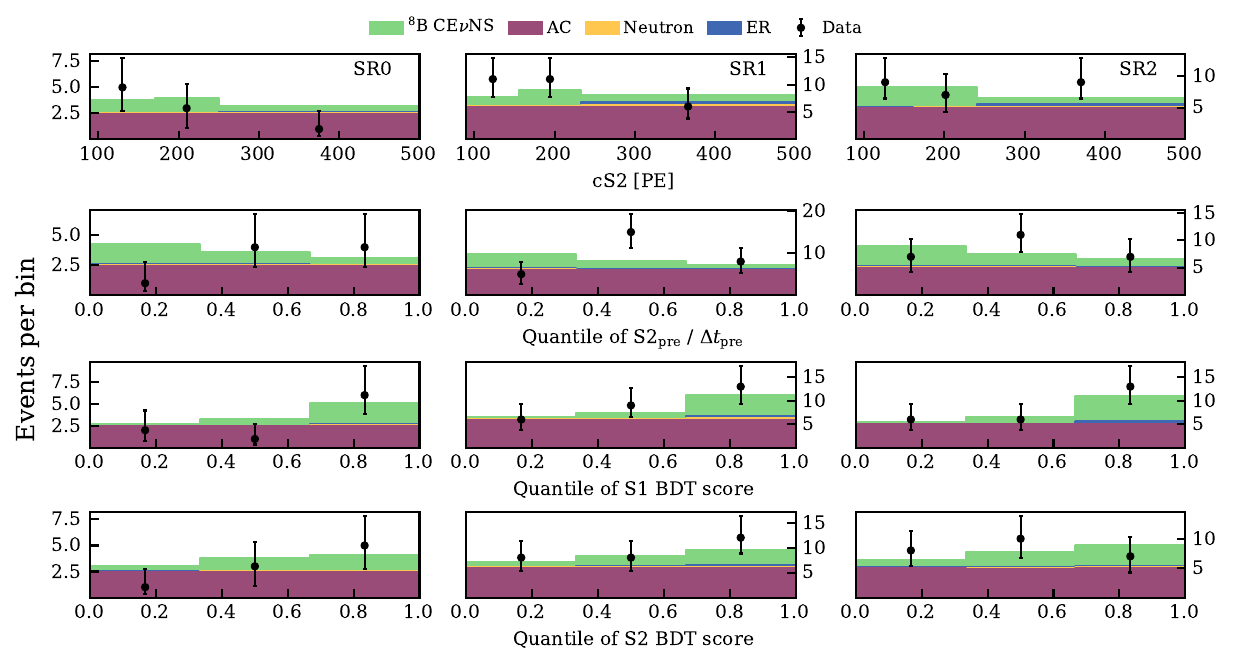}
  \caption{Data and best-fit model projections in the analysis observables for SR0, SR1, and SR2\,(left to right). Black markers indicate the observed event counts with Poisson uncertainties. The stacked histograms show the fitted components, with the \beight~\cevns signal in light green and the backgrounds from AC\,(purple), ER\,(blue), and neutrons\,(yellow).}
  \label{fig:result_breakdown}
\end{figure*}
\twocolumngrid

\appendixsection{Separate SRs best-fit results}
The observed event distributions in SR0, SR1, and SR2, together with the corresponding best-fit model projections in each analysis dimension, are shown separately in \fref{fig:result_breakdown}.

\begin{figure}[!t]
    \centering
    \includegraphics[width=1.0\columnwidth,left]{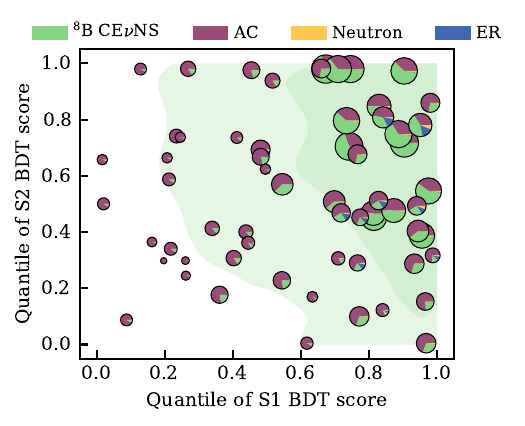}
    \caption{Pie-chart representation of the dataset in the S1 and S2 BDT score plane. All data points are represented as pie charts indicating the fraction of the likelihood of the nominal model, evaluated at the data point, and colored the same as \fref{fig:result_breakdown}. The size of each pie chart scales with the expected \beight signal at the corresponding position. The dark and light shaded contours show the $1\sigma$ and $2\sigma$ expectations of the \beight signal.}
    \label{fig:bdt_score}
\end{figure}

\appendixsection{Event distribution in BDT score}
The distribution of the observed events in the S1 and S2 BDT score plane is shown in \fref{fig:bdt_score}.

\begin{figure}[!h]
    \centering
    \includegraphics[width=1.0\columnwidth,left]{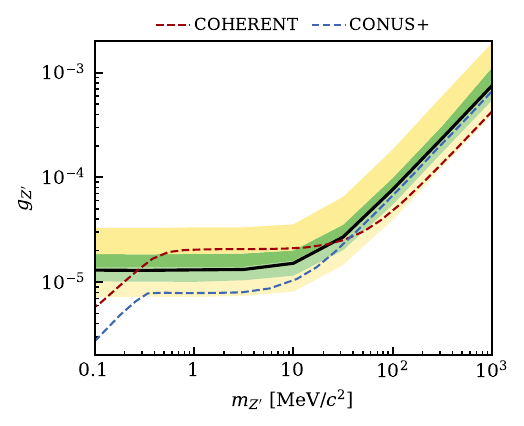}
    \caption{New mediator coupling upper limit at 90\% CL with 1\,$\sigma$\,(green) and 2\,$\sigma$\,(yellow) sensitivity band. Results based on COHERENT\,(red) and CONUS+\,(blue) data are shown for comparison~\cite{AtzoriCorona:2025ygn}.}
    \label{fig:light_mediator_cl}
\end{figure}

\begin{figure}[!t]
    \centering
    \includegraphics[width=1.0\columnwidth,left]{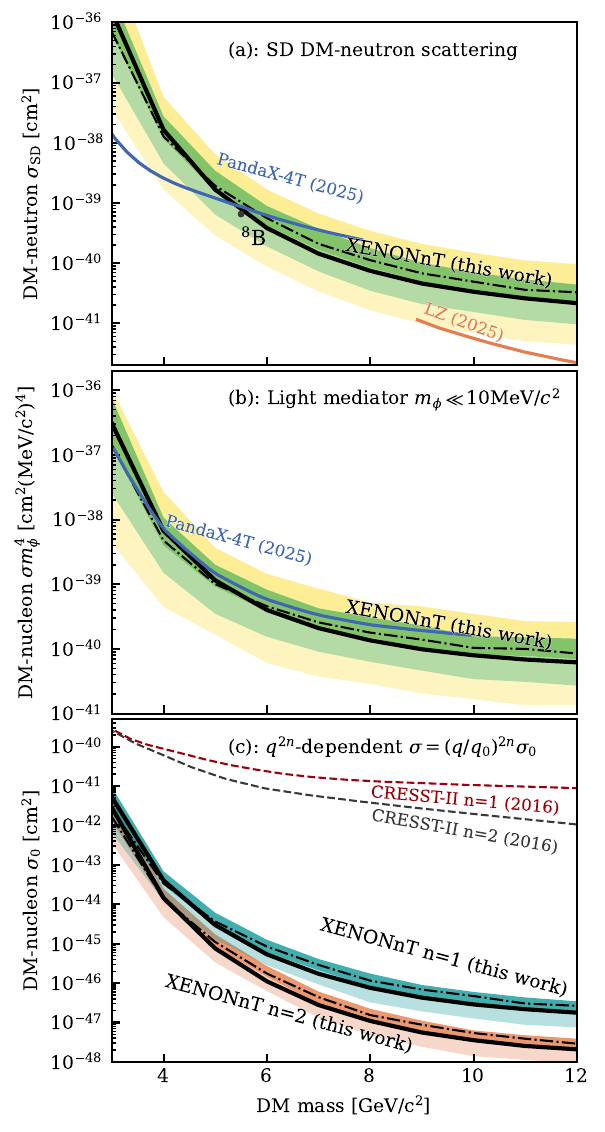}
    \caption{Upper limits on other DM interactions at 90\% CL. The green and yellow bands, as well as the black solid and dash-dotted lines, have the same meanings as in \fref{fig:light_dark_matter_limits_wimp_si}. Results from other experiments~\cite{PandaX:2025rrz, LZ:2024zvo, Angloher:2016jsl} are also shown. (a) SD DM--neutron interaction limit using median nuclear form factors from Ref.~\cite{Klos:2013rwa}; the \beight~\cevns equivalent cross section for a $\SI{5.5}{\gev}/c^2$ WIMP is $6.5\times10^{-40}$\,\unit{cm^2}. (b) Constraints on SI light mediators ($m_\phi \ll \SI{20}{\mev}/c^2$), where the differential rate scales as $\sigma m_\phi^4$. (c) Limits on MDDM interactions for momentum dependence $n=1$ and $n=2$. The teal and orange-filled bands represent the $\pm1\sigma$ sensitivity expectations for $n=1$ and $n=2$, respectively.}
    \label{fig:light_dark_matter_limits_jointed}
\end{figure}

\appendixsection{Additional results on LDM}
The additional models considered here include spin-dependent\,(SD) DM--neutron interactions~\cite{Menendez:2012tm} with an identical mediator as for the SI DM model, and SI interactions mediated by a light scalar or vector mediator with mass $m_\phi \ll q_0/c \equiv \SI{20}{\mev}/c^2$ (SI-LM)~\cite{Fornengo:2011sz,DelNobile:2015uua,XENON:2019gfn}. In the SI-LM case, the differential rate scales as $\sigma m_\phi^4/q^4$, where $q$ is the momentum transfer and $\sigma$ is the DM--nucleon cross section~\cite{PandaX:2022xqx}. In addition, a momentum-dependent dark matter\,(MDDM) scenario~\cite{Chang:2009yt} is examined, characterized by a modified SI interaction cross section of the form $\sigma_{\chi N} = (q/q_0)^{2n}\sigma_0$, with $n \in \{1,2\}$ and $\sigma_0$ the reference DM--nucleon cross section. Upper limits on the DM--nucleon interaction cross section at 90\% CL are shown in \fref{fig:light_dark_matter_limits_jointed}.

\appendixsection{Additional results on BSM neutrino physics}
Several BSM neutrino scenarios are also examined using the same dataset, exploiting the spectral and rate sensitivity to low-energy \cevns events. A new vector mediator $Z'$ that couples to SM leptons and quarks is considered. Assuming no neutrino flavor conversion and a universal coupling strength, the $Z'$ modifies the effective vector coupling as $G_V \propto (g_V^p + 3\epsilon_{ll}^{fV}) Z + (g_V^n + 3\epsilon_{ll}^{fV}) N$, where the additional term $\epsilon_{ll}^{fV} = g_{Z'}^2/[\sqrt{2}G_F(|\vec{q}|^2 + m_{Z'}^2)]$ introduces a momentum-transfer dependence through $|\vec{q}|^2$, with $m_{Z'}$ and $g_{Z'}$ the mediator mass and gauge coupling~\cite{AtzoriCorona:2025ygn,Coloma:2023ixt,Bertuzzo:2021opb}. This enhances the cross section at low $m_{Z'}$ and small $|\vec{q}|$. A flavor-universal neutrino charge radius $\langle r^2_{\nu l} \rangle$ is also considered. This effect modifies the proton vector coupling as $g_V^p \rightarrow g_V^p - \frac{\sqrt{2}\pi\alpha}{3G_F}\langle r^2_{\nu l} \rangle$~\cite{Giunti:2014ixa,AtzoriCorona:2025ygn}, shifting the overall normalization of the proton vector coupling without altering the spectral shape of the NR distribution. When deriving constraints on BSM neutrino interactions, the LDM contribution is set to zero. A 90\% CL limit is set on the coupling constant $g_{Z'}$ as a function of the mediator mass $m_{Z'}$, as presented in \fref{fig:light_mediator_cl}. A 90\% CL limit on the flavor-universal neutrino charge radius $\langle r^2_{\nu l} \rangle$ is obtained as $([-72, -48] \cup [-10, 14])\times10^{-32}\mathrm{cm}^2$, comparable to previous results~\cite{AtzoriCorona:2025ygn}.

\end{document}